# Ferroelectric ε-WO₃ Nanoparticles and Its Bipolaron Driven Opto-electronic Properties at Room Temperature

Mohammad M. Rahaman[1], Jose Flores[2], Mohamed Y. Noor[1], Md Mohsinur R. Adnan[3], Alex Blackston[1], Enam Chowdhury[1], Roberto C. Myers[1,4], Michael Newburger[2], and Pelagia–Irene Gouma[1,5]*

[1]Department of Materials Science and Engineering, The Ohio State University; Columbus, 43210, USA.
[2]Air Force Research Laboratory, Wright-Patterson Air Force; Dayton, 45433, USA.
[3]Department of Electrical and Electronic Engineering, Bangladesh University of Engineering and Technology, 1205, Bangladesh.
[4]Department of Electrical and Computer Engineering, The Ohio State University; Columbus, 43210, USA.
[5]Department of Mechanical and Aerospace Engineering, The Ohio State University; Columbus, 43210, USA.

*Corresponding Email: gouma.2@osu.edu

## Abstract

A unique polymorph of binary tungsten trioxide, the epsilon phase (ε-WO₃), has non-centrosymmetric ferroelectric structure, typically stable below −43ºC in bulk. We have stabilized the ε-WO₃ phase at room temperature (RT) as nanostructured powders via flame spray pyrolysis (FSP) synthesis. These nanopowders are drop cast into uniform thin films to enable RT measurement of ferroelectric and optoelectronic properties. We report ferroelectric hysteresis, nanoscale domains, and dipole switching measured via Piezo-response force microscopy (PFM). The ε-WO₃ films also display optical second harmonic generation (SHG) and anticlockwise ferroelectric butterfly capacitance versus voltage hysteresis, further demonstrating the ferroelectric nature of ε-WO₃. Remarkably, ε-WO₃ shows ferroelectric polarization responses to optical stimuli and forms bipolaron at RT− a spin-zero quasiparticle previously found only in cryogenic temperatures. The bipolaron formation and its interaction with electro-optical stimuli results in a single layer solid-state blue coloration, a ferrochromic effect. A mechanism of the ferrochormic effect are discussed. In summary, ε-WO₃ appears to be a ferroelectric with the simplest structure, forming bosonic spin-zero bipolarons, and its dipoles respond to opto-electrical signals; therefore, this material holds significant promise for transforming the field of optoelectronics.

## 1. Introduction

Tungsten trioxide ($WO_3$) is a binary polymorphic metal-oxide with advanced functionalities, and diverse applications from energy to electronics[1-6]. It typically exhibits the simplest form of the perovskite $ABO_3$ structure, notably lacking the A-site cation. It undergoes a series of crystallographic phase transformations- from a high temperature (above 900ºC) cubic phase to a low temperature (below −43ºC) monoclinic ε-phase (space group $P_c$)[7-9]. These transformations are primarily driven by the displacement of $W^{6+}$ and $O^{2-}$ ions, as well as the tilting and rotation of $WO_6$ octahedra [10]. The low temperature stable polymorph, the ε-phase, specifically, has garnered significant interest because of its acentric structure which theoretically polarizes the ions[10].

The interplay of competitive dipole moments in $WO_6$ octahedra potentially make the ε-phase simplest known (binary) ferroelectric material [10]. The exploration on ferroelectric properties in ε-WO₃ began in 1950's when Matthias et al. from Bell telephone laboratories first proposed the existence of ferroelectricity in the material, based on unusual low temperature electrical behavior [11, 12]. Since then, extensive efforts have been devoted on understanding and establishing the ferroelectric nature of ε-WO₃ [13, 14]. A significant milestone was achieved in 2006, when Hirose et al. employed single crystal of $WO_3$ and a modified Sawyer-Tower circuit to demonstrate, for the first time, an electrical hysteresis loop of dielectric displacement vs electric field at **−100°C**[15].

Previously, ε-phase was stabilized at RT and above revealing evidence of ferroelectric hysteresis in sintered $WO_3$ pellets of about 200 nm grain size at −190ºC [3, 16, 17]. This paper reports the ferroelectric behavior of ε-$WO_3$ nanoparticles in the form of a nanocrystalline thin film, the ferroelectric dipoles' interaction with light, and a mechanism of ferrochromic behavior of ε-$WO_3$.

## 2. Experimental Details

### 2.1 Materials

Here, the reported ε-$WO_3$ was synthesized via flame spray pyrolysis (FSP) technique in the form of powder. The synthesis process is discussed in ref.[3, 16]. The synthesized powder predominantly consists of γ-phase, with small fraction of ε-phase [16]. The ε-phase fraction is enhanced through a separation process utilizing a non-uniform electric field generated by an electrospinning set-up. Details of this method are provided in supplementary materials [Figure S1(A, B) and Movie S1]. When the mixed samples are exposed to a non-uniform 2 kV/cm electric field, the ε-$WO_3$ nanoparticles interact with the field via their intrinsic ferroelectric dipoles, exhibiting translational motion toward the opposite side of the beaker. This approach presents a novel solid-state technique for separating ε-phase from a multiphase system to enable characterization of its ferroelectric and optoelectronic response.

The purified powder of ε-$WO_3$ is used to deposit films on conducting ITO substrate and ITO interdigitated-electrode via drop cast technique. Making uniform and defect-free film from powder over a large conducting area without altering the crystallographic structure, stoichiometry and particle size for properties measurement and device fabrication is a challenge, since conventional physical and chemical techniques change the crystallographic phase of the materials, and lead to non-uniform, porous and cracked film formation, respectively [18-21]. In this study, uniform films of ε-WO$_3$ on conducting indium tin oxide (ITO) substrates (Sigma-Aldrich) from powder are prepared using a controlled drop cast technique. For this, 0.004 mg/μL water solution of ε-WO$_3$ powder is prepared inside of 0.6 mL micro-centrifuge tube. The solution is then ultrasonicated for 30 minutes in a Branson 2510 sonicator. Afterwards, 50 μL of the solution is drop-cast on clean ITO substrate, while the substrate is kept on a flat hot plate at 80$^o$C. The superhydrophobic nature of the ITO substrate leads to circulatory flow within the droplet that counteracts the outward capillary flow, preventing the coffee ring formation [22]. As a result, the contact line retracts evenly as the solvent evaporates (shown in Movie S2), resulting in uniform film formation. Another film of ε-WO$_3$ is also deposited on ITO interdigitated electrodes following the same procedure as described in reference [6]. The films are referred to as F-1 and F-2 throughout the manuscript respectively.

## 3. Results and Discussion

### 3.1 Ferroelectric nature of e-WO3 nanoparticles in thin film at RT

The XRD of the purified ε-WO$_3$ powder contains X% ε-phase, with Y% γ-phase, shown in Figure S2(A). The thickness of the film is 3.5 μm as presented in Figure S2B. High resolution surface topography of the film is assessed using Atomic Force Microscopic (AFM) images within 500 x 500 nm$^2$ area and presented in Figure 1(A). The film surface is covered with randomly distributed polycrystalline grains ranging from 8 to 30 nm. The rms surface roughness of the film is 5.61 nm.

The peak to valley distance is approximately 30 nm as measured from height profile of Figure 1(B). These results indicate good uniformity of the drop-casted film.

Phase-resolved PFM is used to explore the local ferroelectric switching properties of the films. Here, amplitude images provide the surface topography, while phase contrast resolves local piezoelectric responses. The contrast observed in Figure 1(C) and Figure 1(E) show PFM responses in the out of plane (OP) and in-plane (IP), orientations, respectively. The OP PFM response of $\varepsilon$-$WO_3$ nanocrystals exhibit random orientation within the phase of $-80^{\circ}$ to $+80^{\circ}$. The random orientation of these nanocrstals might have originated from polycrystalline grains of the as-deposited thin film. The comparison of surface morphology and the OP PFM image in Figure 1(A) and 1(C) indicates that individual nanocrystals are giving rise to ferroelectric nanodomains. The phase of OP nanodomains are nearly constant within the grains and shift sharply near grain boundaries, as observed in the phase profile in Figure 1(D) (plotted along line-1 of Figure 1(C)). In contrast, the phase of most IP nanodomains is oriented at $-20^{\circ}$. However, IP domains ranging from $-60^{\circ}$ to $180^{\circ}$ are also observed located near the grain boundaries. The nanodomains of $\varepsilon$-$WO_3$ were further investigated by sampling a large area ($1 \times 1 \ \mu m^2$) and the results are presented in supplementary Figure S4.

Testing for ferroelectric response by PFM is carried out by examining hysteretic behavior of the phase contrast by applying local electric field within $\pm 28.5$ kV/cm, see Figure 2. A clear hysteresis is observed at a local position. The phase of the nanodomains switches from $-10^{\circ}$ to $-90^{\circ}$, leading to a well-defined symmetrical hysteresis loop as shown in Figure 2(A). In Figure 2(B), the amplitude variation of the nanodomains in response to the electric field reveals butterfly hysteresis with two inflection points at around $\pm 3.3$ kV/cm, which is the coercive field ($Ec$) of the material.

The calculated $Ec$ value is close to that reported by Owen at el. in his dissertation, conducted on sintered pellet of $WO_3$ at $-190^{\circ}C$, where epsilon phase is known to be stable[17]. However, PFM measurement alone does not comprehensively prove ferroelectricity of $\varepsilon$-$WO_3$ since electrostriction and electrostatic effects in between PFM tip and film's surface might lead to PFM hysteresis. To rule out these possibilities, we have performed second harmonic generation of $\varepsilon$-$WO_3$− a tool provides discerning evidence on non-centrosymmetric structure of the materials, and capacitance vs voltage measurement showing full 180-degree (or pi) phase switch between domains.

The asymmetric structure of $\varepsilon$-$WO_3$, domain orientations and switching, the second harmonic generation (SHG) is employed on F-2 using the rotational anisotropy harmonic generation (RAHG) method. The sample is excited with 1030 nm (i.e. 1.2 eV sub-band gap excitation since $WO_3$ bandgap is 3.2 eV[https://doi.org/10.1016/j.jlumin.2017.11.053; https://doi.org/10.1016/j.solener.2014.11.014), 77 fs ultrafast laser source in transmission mode to avoid collective surface SHG contributions. As shown in Figure 3(A), $\varepsilon$-$WO_3$ shows remarkable SHG signal. The SHG intensity generated from $\varepsilon$-$WO_3$ film is ~15 times higher than the background ITO interdigitated electrode substrate. This strong SHG response confirms that the observed nonlinear optical behavior originates from the non-centrosymmetric $\varepsilon$-$WO_3$ phase. It has been reported that SHG intensity scales with the square of the net polarization within the probed volume. Therefore, any change in net polarization or domain configuration due to external electric stimuli is reflected in the SHG response[23, 24]. In Figure 3(B), the SHG intensity is shown to vary with the applied voltage, indicating that the multidomain configuration of $\varepsilon$-$WO_3$, as previously observed through PFM, evolves in response to the external bias. The SHG signal initially increases from 0 V and reaches a switching point near 2.3 V, which closely corresponds to the coercive voltage identified from the capacitance–voltage (C–V) curve in Figure 3(E). One plausible explanation for this voltage-dependent SHG behavior in multi-domain system is that, as the applied voltage increases up to the coercive threshold, ferroelectric domain switching occurs while leaving behind some residual non-inverted domains[25]. These remaining domains may constructively contribute to the overall SHG signal, leading to the observed intensity increase. Beyond the coercive threshold, as domains switch in opposite direction, the SHG contributions from the inverted and non-inverted domains may interfere destructively, resulting in a decrease in signal intensity. (Note: similarities with van der Waals-layered ferroelectric $CuInP_2S_6$, which exhibits enhanced SHG intensity and increase of in-plane polarization orientation below the coercive threshold when the electric field is applied parallel to the in-plane direction[26])

Additionally, the voltage-dependent SHG behavior observed in ε-$WO_3$ aligns with previous studies conducted on initially polydomain systems[27]. While further investigation is required to determine whether residual non-switched domains persist or if domain switching follows a needle-like morphology, the present observations strongly indicate that the ferroelectric domains in ε-$WO_3$ are switchable under applied bias, a consistent conclusion with both the PFM and C–V measurements.

Furthermore, the SHG provides important information on domain orientation of ferroelectric ε-$WO_3$. The polar plots of SHG intensity for ε-$WO_3$ are shown in Figure 3(C, D), where the collected SHG is in parallel and perpendicular orientation respectively with respect to the fundamental beam as shown in Figure S5. The important observations on the polar plots are: presence of anisotropic two-lobed SHG patterns having maxima around 50° and 230° in both configurations. The orientation of ferroelectric dipoles observed in the polar study closely resembles OP nanodomains orientation shown in the PFM study in Figure (1C). The polar plot pattern of the nanocrystalline film is similar to that of a uniaxial crystal [28] which indicates that SHG may originate from type-I. From Figure 3C, when P1 and P2 are in perpendicular configuration we see a dominant two lobe pattern. In parallel configuration in Figure 3D, SHG intensity is diminished significantly. The lower intensity might be the reminiscent signal from interface/substrate.

SHG is a sensitive technique for probing the polarity of ferroelectric domains; however, it does not directly address the possibility of electrostriction or electrostatic artifacts observed in PFM studies, which can mimic ferroelectric-like hysteresis. Thus, to confirm definitely that the hysteresis observed in PFM arises from the ferroelectric domains of $\varepsilon$-WO$_3$ and that these domains show 180° polarization reversal, we have performed capacitance–voltage (C–V) measurements at $10^6$ Hz on the F-2 film. The measurements were initially conducted under dark conditions, revealing an anticlockwise butterfly-shaped hysteresis loop, as shown in Figure (3E). This response provides strong evidence that our FSP synthesized $\varepsilon$-WO$_3$ exhibits room-temperature ferroelectricity. Additionally, the appearance of two distinct peaks around ±2.5 V indicates 180° domain switching. Together, the C–V results not only confirm the existence of $\varepsilon$-phase but also establish its ferroelectric nature—resolving scientific debate that has persisted for a better part of a century.

## 3.2 Ferro-chromism in $\varepsilon$-WO3

A remarkable characteristic of the $\varepsilon$-WO$_3$ is its ability to exhibit blue coloration in solid-state between interdigitated electrodes upon application of electric bias without any presence of electrolytes or ions —observable under natural light and notably occurring even under vacuum or inert gas environments, please see the supplementary Figure S6 for details. This blue coloration consistently initiates from the negative terminal of the interdigitated electrodes and propagates toward the positive side. The phenomenon was systematically observed under various conditions, including ambient air, inside of an argon-filled glovebox, and under vacuum [6]. In all cases, the blue coloration was triggered exclusively by electrical bias only. This phenomenon makes ε-WO₃ suitable for single layer solid state electrochromic display, avoiding the requirement of $H^+$, $Li^+$, or $Na^+$ ions insertion as it was always needed for gamma or amorphous $WO_3$[29-31]. Previous studies have suggested that this effect may arise from the interaction of light with the electrically induced alignment of ferroelectric nanodomains in ε-WO₃[6]. Here, we have carried out comprehensive opto-electrical studies to elucidate the underlying cause of the coloration of ε-WO₃, and to explore the relationship between ferroelectric domain orientations and this unique solid-state coloration phenomenon.

The asymmetric $WO_6$ octahedral framework provides a favorable structural support for the formation of polarons and bipolarons [32]. In a pioneering work during the 1980s, Schirmer et al. investigated polaronic transport in ε-WO₃ at cryogenic temperatures, revealing the formation of a small to medium bi-polaron with an energy of 1.1 eV under dark conditions at 20 K. Upon exposure to light, the bi-polaron dissociated into mobile polarons [33]. Subsequent studies have demonstrated that the optical absorption features and charge transport in ε-WO₃ are closely linked to polaronic and bi-polaronic mechanisms, often associated with oxygen vacancies [33-35]. However, recent theoretical studies suggest that oxygen vacancies are not necessary for polaron formation. Instead, the emergence of ferroelectricity and it's corresponding polar zone-center mode in $WO_3$ are the key driver for polaronic behavior upon insertion of external electrons into the lattice [36].

We have proven that bi-polaron forms in ferroelectric ε-WO$_3$ at room temperature under electrical stimuli. The formation and dissociation of bi-polaron by the interaction with light creates conduction pathways in between the electrodes, resulting in blue coloration in ε-WO$_3$.

In Figure (3E), a ferroelectric anticlockwise butterfly loop of ε-WO$_3$ is observed under dark conditions. However, the ferroelectric hysteresis loop disappears and charged trapped clockwise hysteresis appears when the measurement is performed in presence of light as shown in Figure (3F), indicating the ferroelectric domains interacting with light. The value of capacitance under light decreases, which strongly demonstrates the increase of conduction from charge trapping, resulting in blue coloration in the material. This poses a critical question: does the ferroelectric domains interaction with electro-optical stimuli play a role in the coloration process of ε-WO$_3$?

To address the question and identify the interaction region of the charge trapping, transmittance spectra are collected within 400 to 1100 nm wavelength at different voltages. The values of transmittance (T) are used to calculate optical density (OD) by utilizing the following formula, OD = $-\log_{10}$(T)[37], which later utilized to calculate $\Delta O$D for further analysis. The "Optical density" refers to the material's absorbance of light. The term $\Delta$OD represents the change in optical density (absorbance), where an increase or decrease in $\Delta$OD corresponds to a respective increase or decrease in light absorption by the material. As shown in Figure 4(A), the $\Delta$*(Optical Density)* vs *Wavelength* plot at 20 V reveals a broad absorption feature peaked at 1.2 eV (~1030 nm). This absorption peak is not attributed to impure γ- or δ-WO$_3$ as γ-WO$_3$ exhibits polaronic absorption around 0.4 eV, and δ-WO$_3$ lacks polaronic activity[34, 36, 38]. Instead, the broad peak is attributed to ε-WO$_3$ and represents bi-polaronic absorption. The formation of bi-polaron in ε-WO$_3$ room temperature is consistent with earlier findings, where bi-polaron in ε-WO$_3$ was reported at 20 K under illumination confirmed via electron spin resonance[33, 35]. In addition, our finding satisfies the bi-polaron stability condition i.e. $|E_b| < 2E_s$, where $E_b$ is the peak optical absorption energy of bi-polaron which is 1.2 eV in the present work, and $E_s$ is absorption energy of polaron reported to be 0.7 eV [33]. Notably, the lower bound of the bi-polaron absorption begins at approximately 620 nm (2.0 eV), indicating that red-light (620 - 750 nm) is absorbed from the visible spectrum, resulting in the observed blue coloration of $\varepsilon$-WO$_3$.

Figure 4(B) establishes a correlation between $\Delta$(Optical Density) and SHG intensity with applied voltage in presence of optical stimuli at 1.2 eV. Near the coercive voltage, $\Delta$(Optical Density) is minimum. As the applied voltage increases beyond the coercive threshold, the value of $\Delta$(Optical Density) increases. This rise of optical absorption is an indication of increasing bi-polaronic activity. At the same time, the corresponding value of SHG intensity decreases, and the ferroelectric domains start to demonstrate $180^{\circ}$ switching as observed from C vs V measurement in Figure 3(E). This suggests a correlation between the bi-polaronic activity with ferroelectric domain orientations. When the orientations of ferroelectric domains increase with the electric field, subsequent bi-polaron formation and dissociation also increases.In summary, when a potential is applied in between the electrodes, the ferroelectric dipoles of $\varepsilon$-WO$_3$ get aligned with the field, which leads to 5d orbitals of W$^{6+}$ ions in asymmetric WO$_6$ octahedra to accept electrons from the negative electrode via a Jahn-Teller effect or anti-distortive polaron formation mechanism, altering the dipole moment of W$^{6+}$ and converting W$^{6+}$ into W$^{5+}$[39-41]. According to salje's model, the localized electrons in W$^{5+}$ ion are expected to form a polaron. Afterwards, two neighboring polarons form a bipolaron through the 'inter-site bipolaron model' [34]. The bipolaron then dissociates into mobile polarons by absorbing light of wavelength greater than 620 nm as observed in Figure 4 [34]. Once dissociated, the mobile polarons hop to the neighboring site near the positive electrode. This starts a new cycle of formation and dissociation of bipolaron in the presence of light. The process continues until the electron reaches the positive electrode. Following this process, continuous pathways to transport charge between the electrodes are established under light. As a result, the coloration starts from the edge of negative electrode and propagates towards the positive electrode. The propagation of coloration and $W^{6+} \rightarrow W^{5+}$ transition has been confirmed via in-situ optical microscopy and X-ray photoelectron spectroscopy respectively in our previous work in ref[6].

**Conclusion**

In conclusion, this work presents evidence of ferroelectricity in ε-WO$_3$ nanocrystalline films, it's room temperature bi-polaron formation, and a mechanism on solid-state coloration. The key findings of this study are: (i) the ε-WO$_3$ shows multiple orientations of PFM response indicative of ferroelectric domains. The ferroelectric domains switch under electrical signal, observed via PFM, SHG. Full 180° domain switching is observed in ensemble C vs V measurement. (ii) Ferroelectric dipoles interact with optical signals, supported by C–V data performed in dark and light. (iii) Bi-polaron forms in ε-WO$_3$ at room temperature through which optical interaction occurs. (iv) The bi-polaronic activity correlates with the ferroelectric domain orientations and accounts for the single-layer solid state blue coloration in ε-WO$_3$, the ferrochromic effect. These unique properties of structurally simple and ferroelectric ε-WO$_3$ presents it as a promising pathway for the development of energy-efficient opto-electronic technologies like light driven artificial neuron, ferroelectric memories, high temperature superconductor, or solid-state displays etc. Epsilon-WO$_3$ polymorph is emerging as true binary ferroelectric, and it deserves to be further explored now that it has become available at RT and above.

# CRediT Authorization Contribution Statement

**Mohammad Mahafuzur Rahaman:** Conceptualization, Data Curation, Investigation, Methodology, Software, Formal Analysis, Writing−original draft, Writing− review and editing; **Jose Flores:** Data curation, Investigation, Methodology, Software, Formal Analysis, Validation, Visualization, Writing− review and editing; **Mohamed Yasin Noor:** Data curation, Investigation, Software, Formal Analysis, Validation, Visualization; **Md. Mohsinur R. Adnan:** Data curation, Investigation, Methodology, Software, Formal Analysis, Validation, Visualization; **Alex Blackston:** Data curation, Software, Validation, Visualization; **Enam Chowdhury:** Resources, Funding acquisition, Supervision, Methodology, Investigation, Validation, Writing− review and editing; **Roberto C. Myers:** Conceptualization, Resources, Software, Funding acquisition, Supervision, Methodology, Investigation, Validation, Writing− review and editing; **Michael Newburger:** Resources, Software, Funding acquisition, Supervision, Methodology, Investigation, Validation, Writing− review and editing; **Pelagia-Irene Gouma:** Conceptualization, Project Administration, Fund acquisition, Supervision, Investigation, Methodology, Software, Writing− review and editing.

# Funding

Support for this work was provided by:

Orton Ceramic Foundation/Orton Chair Funds-GF 605515

National Science Foundation (NSF) Award No. IIS2014506

Air Force Office of Scientific Research (AFOSR) Award No. FA955023RXCOR02

Air Force Office of Scientific Research (AFOSR) Award No. FA9550-20-1-0278

The Ohio State University College of Engineering (COE) Strategic Research Initiative Grant titled "Next generation Electro-Optics and Nanophotonics (NEON)"

## Declaration of competing interest

The authors declare that they have no known competing financial interests or personal relationships that could have appeared to influence the work reported in this paper.

## Acknowledgement

The authors acknowledge S. Jabin, Electrical and Computer Science Engineering, The Ohio State University for his help in electrical study.

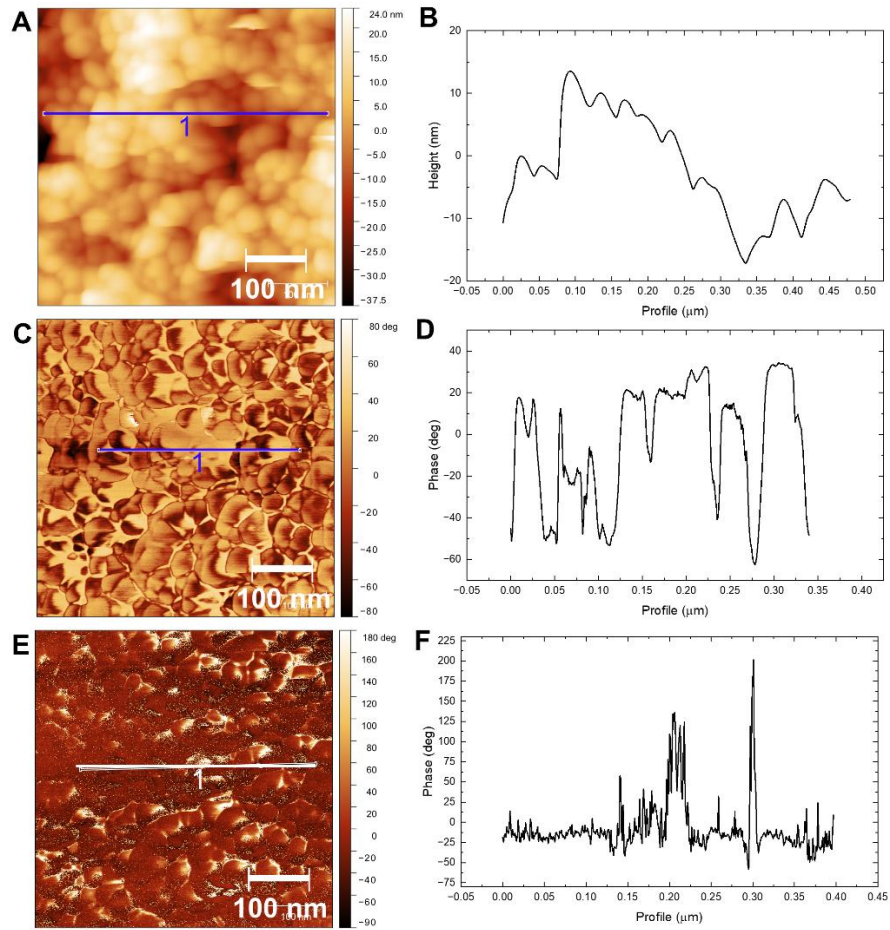

**Figure 1. Surface topography and ferroelectric nanodomains of ε−WO₃. (A, B)** Surface morphology of ε−WO₃ film and height profile measured by AFM. **(C, D)** Out of plane ferroelectric nanodomains of ε−WO₃ and their phase profile. **(E, F)** In-plane ferroelectric nanodomains of ε−WO₃ and their phase profile. [this study is performed at RT using ε−WO₃ film deposited on ITO substrate].

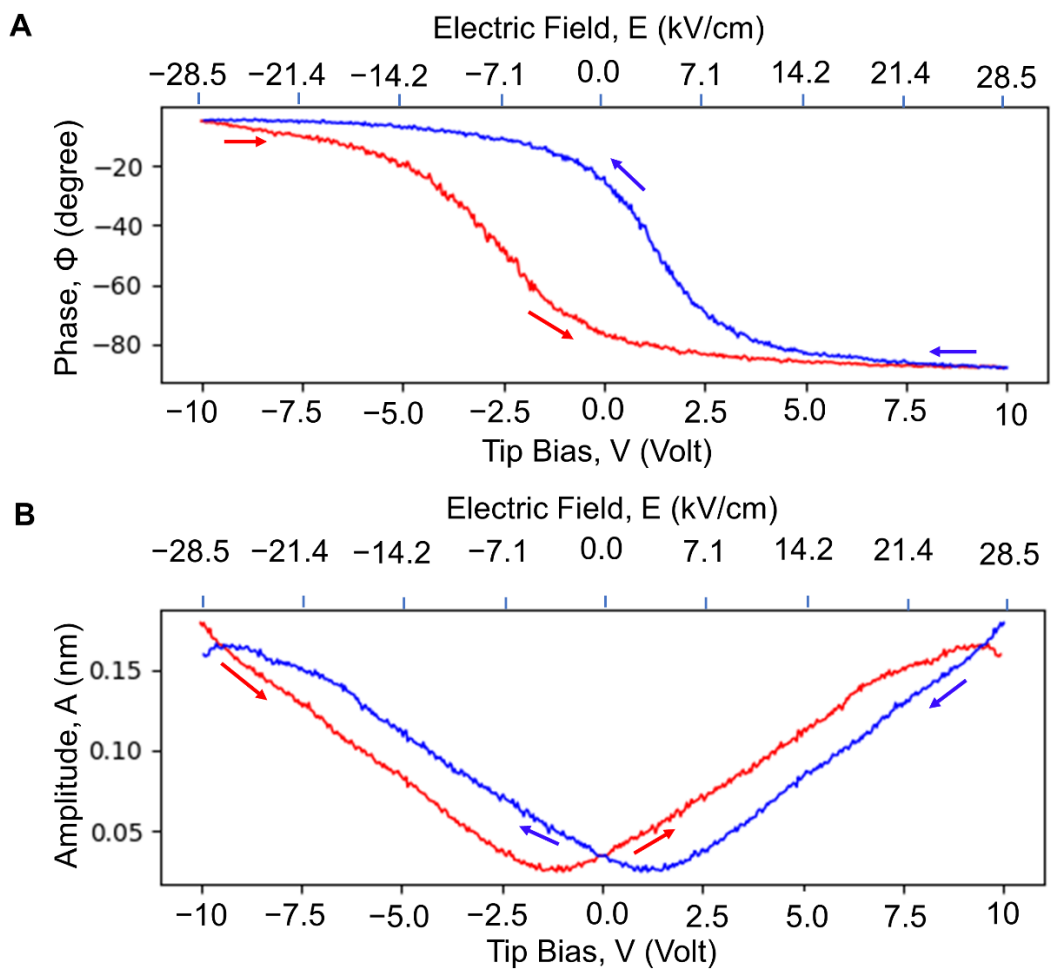

**Figure 2. Ferroelectric hysteresis of ε−WO₃ at room temperature.** (**A**) Phase switching of the nanodomains of ε−WO₃. and (**B**) Butterfly type hysteresis of amplitude variation of the nanodomains, within ±28.5 kV/cm electric field studied by PFM.

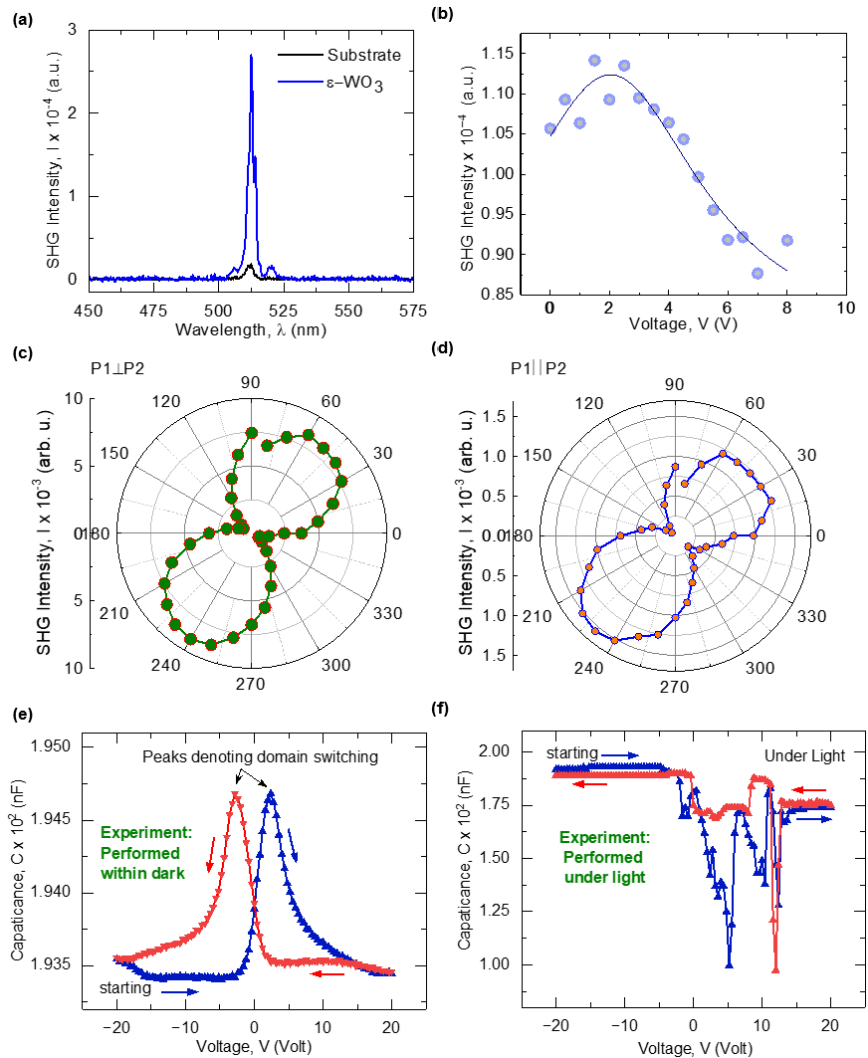

**Figure 3. Opto-electrical studies of ferroelectric domains of ε-WO₃ (a)** Comparison of SHG signal of ε-WO₃ with substrate, **(b)** Voltage dependent SHG intensity, **(c, d)** Polarization dependent study of ε-WO₃ without bias in parallel and perpendicular configurations, respectively [Perpendicular direction from negative to positive electrode in interdigitated electrodes is considered as 0° here. The sample was biased upto ~20 V before our SHG measurements]. **(e, f)** C vs V studies illustrate the ferroelectric anticlockwise butterfly loop in dark, while clockwise charge trapping loop under light conditions [the C vs V measurement was performed at 1 MHz frequency and started from -20 V].

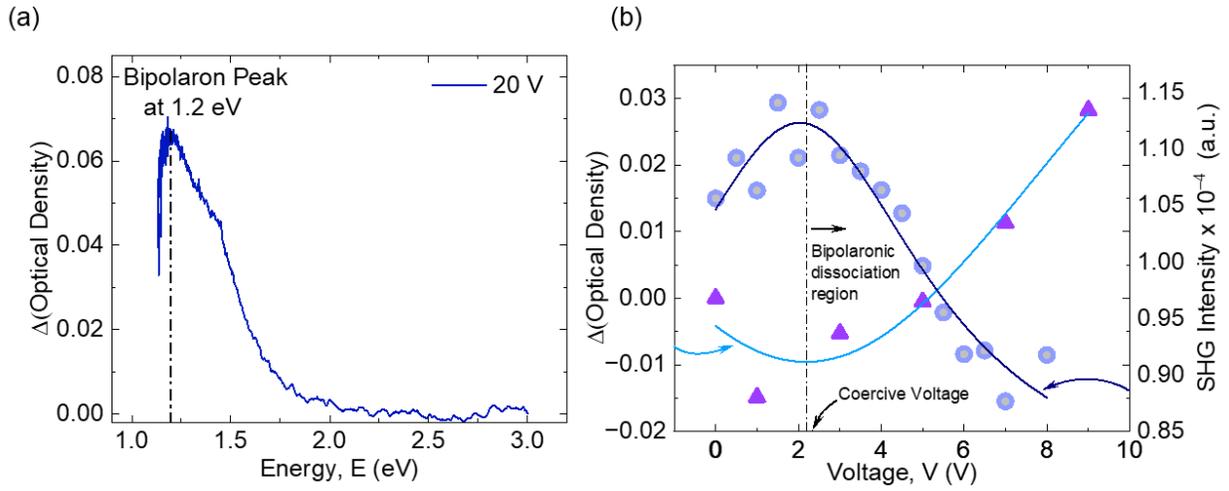

**Figure 4: Optical studies on bipolaron;** (a) RT bipolaron signature in ε-WO$_3$ peaked at 1.2 eV, (b) Change in optical density vs voltage for maximum bipolaron dissociation energy at 1.2 eV and its correlation with domain switching.